\documentstyle[11pt]{article}

\def\be{\begin{equation}}
\def\ee{\end{equation}}
\def\bea{\begin{eqnarray}}
\def\eea{\end{eqnarray}}

\gdef\makemath#1{\ifmmode #1 \else $ #1 $\fi}
\newcommand{\ignore}[1]{}

\newwrite\noteFile%
\immediate\openout\noteFile=\jobname.notes%

\def\saveNote#1{%
\immediate\write\noteFile{Page \expandafter\thepage; }%
\immediate\write\noteFile{\expandafter#1}%
\immediate\write\noteFile{}%
}

\def\authNote#1{}

\def\pair#1{\langle #1 \rangle }

\def\set#1{\{#1\}}

\def\intersect{\cap}

\renewcommand{\Re}{{\rm Re}}

\def\QED{\hfill\fbox{\phantom{.}}}

\newenvironment{mytheorem}[1]%
{\xdef\kindOfTheorem{#1}\begin{\kindOfTheorem}}%
{\end{\kindOfTheorem}}%

{\par\medskip\noindent{\bf #1}\begin{it}}%
{\end{it}\par\medskip}

\newcommand{\strings}{\makemath{\{ 0,1 \}^{*}}}

\def\compclassfont#1{{\sf{#1}}}

\def\PH{\compclassfont{PH}}
\def\NP{\compclassfont{NP}}

\def\P{\compclassfont{P}}

\def\FP{\compclassfont{FP}}
\def\PP{\compclassfont{PP}}

\def\BPP{\compclassfont{BPP}}

\def\NQP{\compclassfont{NQP}}
\def\EQP{\compclassfont{EQP}}
\def\BQP{\compclassfont{BQP}}
\def\GapP{\compclassfont{GapP}}
\def\CequalsP{\mbox{\compclassfont{C$_=$P}}}
\def\coCequalsP{\mbox{\compclassfont{coC}$_=$\compclassfont{P}}}

\def\sharpP{\compclassfont{\#P}}

\def\Count#1#2{|#2|_{#1}}
\newcommand{\ket}[1]{{|{#1} \rangle}}

\newcommand{\braket}[2]{{\langle {#1} \mid {#2} \rangle}}
\newcommand{\tuple}[1]{{\langle {#1} \rangle}}
\newcommand{\strs}{{\Sigma^*}}
\newcommand{\nums}{{\bf N}}

\newcommand{\rats}{{\bf Q}}
\newcommand{\reals}{{\bf R}}
\newcommand{\complexes}{{\bf C}}
\newcommand{\algebraics}{{\overline{\bf Q}}}

\newcommand{\bfvec}[1]{{\bf{#1}}}
\renewcommand{\bfvec}[1]{{\vec{#1}}}

\def\bfvec#1{\mathchoice{\mbox{\boldmath$\displaystyle\bf#1$}}
{\mbox{\boldmath$\textstyle\bf#1$}}
{\mbox{\boldmath$\scriptstyle\bf#1$}}
{\mbox{\boldmath$\scriptscriptstyle\bf#1$}}}

\begin{document}

\title{Determining Acceptance Possibility for a Quantum Computation 
is Hard for the Polynomial Hierarchy}

\author{
Stephen Fenner~\thanks{Computer Science Department, University of
South Carolina, Columbia, SC 29208 (on leave from the University of
Southern Maine).  E-mail: fenner@cs.sc.edu.
Supported in part by the NSF under grant and CCR 95-01794.}\\
\normalsize
University of South Carolina
\and
Frederic Green~\thanks{%
Department of Mathematics and Computer Science, Clark University, Worcester, MA
01610.  E-mail: fgreen@black.clarku.edu.  
}\\
\normalsize
Clark University 
\and
Steven Homer~\thanks{Computer Science Department, Boston University,
  Boston, MA 02215. E-mail: homer@cs.bu.edu. Supported in part by the NSF
under grant NSF-CCR-9400229.}\\
\normalsize
Boston University
\and
Randall Pruim~\thanks{Department of Mathematics and Statistics, 
Calvin College, Grand Rapids, MI  49546.
E-mail: rpruim@calvin.edu. 
This work was done while visiting the Computer Science Department at
Boston University.}\\
\normalsize
Calvin College
}

\maketitle

\begin{abstract}
	It is shown that determining whether a quantum computation has a
non-zero probability of accepting is at least as hard as the
polynomial time hierarchy.  This hardness result also applies to
determining in general whether a given quantum basis state appears
with nonzero amplitude in a superposition, or whether a given quantum
bit has positive expectation value at the end of a quantum
computation.  This result is achieved by 
showing that the complexity class $\NQP$ of
Adleman, Demarrais, and Huang \cite{ADH97}, a quantum analog of $\NP$,
is equal to the counting class $\coCequalsP$.

\end{abstract}

\section{Introduction}

This decade has seen renewed interest and great activity in quantum
computing.
This interest has been spurred by the clear formal definition of the
quantum computing model and by the surprising discovery that some
important computational problems which may be classically
infeasible are feasible using quantum computers.  One central result
is Shor's bounded-error polynomial-time algorithms for discrete
logarithm and for integer factoring on both a quantum Turing machine
\cite{Shor:QuantumFactoringFOCS} and (equivalently) quantum circuits
\cite{Shor:QuantumFactoring}.  This opens the possibility that if such
machines can be constructed, or effectively simulated, then one can
rapidly factor large integers and compromise a good deal of modern
cryptography.

While the main research focus has been on finding efficient quantum
algorithms for hard problems, attention has also been paid to
determining the strength of quantum computation {\it vis-\`{a}-vis\/}
its classical (probabilistic) counterpart \cite{BB92,BV97}.
In this paper we take a further step in this
direction by proving that testing for non-zero acceptance probability
of a quantum machine is classically an extremely hard
problem.
In fact, we prove that this problem---which we call $QAP$
(``quantum acceptance possibility'') and which is complete for $\NQP$ 
(a quantum analog of $\NP$)---is hard for the polynomial-time hierarchy.  
This is done by showing that $\NQP$ is precisely the exact counting class
\cite{Wagner:C=P} 
$\coCequalsP$:

\begin{mytheorem}{Theorem}
\label{thm:equiv}
$\NQP = \coCequalsP$.
\end{mytheorem}

$\coCequalsP$, in turn, is hard for $\PH$ under randomized
reductions~\cite{Toda:PHinBPParityP,TO:PHinBPGapP}, and may still be
hard even if $\P = \NP$.  Thus

\begin{mytheorem}{cor}
\label{thm:hard}
The problem of determining if the acceptance probability of a quantum 
computation is non-zero ($QAP$) is hard for the polynomial time
hierarchy under polynomial-time randomized reductions.
\end{mytheorem}

We will see in Section~\ref{sec:robustness} that
Theorem~\ref{thm:equiv} is mostly insensitive to the set of transition
amplitudes we allow in our model of quantum computation.  The equation
holds whether we allow arbitrary algebraic numbers as transition
amplitudes (Theorem~\ref{thm:algebraic}) or we restrict transition
amplitudes to be in a small finite set of rational numbers as
described by Adleman, {\em et al.\/} \cite{ADH97}
(Theorem~\ref{thm:rat}).  We will assume throughout the paper that
transcendental amplitudes are not allowed.

The class $\NQP$ was originally defined by Adleman, Demarrais, and
Huang \cite{ADH97}, who showed that $\NQP \subseteq \PP$. The sharper
upper bound $\NQP \subseteq \coCequalsP$ is implicit in their proof
and a recent result of Fortnow and Rogers \cite{FR:BQP}. The main
contribution of this paper is to obtain the {\it lower bound}
$\coCequalsP \subseteq \NQP$. Adleman et al. also asked if $\EQP$ (the
quantum analog of $\P$) and $\NQP$ are the same. Our result implies
that $\EQP = \NQP$ is equivalent to the collapse of the counting
hierarchy (see Section~\ref{sec:main}).

Graph Nonisomorphism \cite{KST} is an example of a problem in
$\coCequalsP$ that is not known to be in $\NP$.
Theorem~\ref{thm:equiv} shows that there is a quantum machine that
takes two graphs as input and accepts with probability zero exactly
when the two graphs are isomorphic.

We prove Theorem~\ref{thm:equiv} and Corollary~\ref{thm:hard} in
Section~\ref{sec:main}.  The proof can be easily adapted to show
hardness of determining whether any given quantum bit must be zero (or
one) with certainty in a quantum computation, or more generally,
whether some given quantum state shows up in a superposition with
nonzero amplitude.  Both of these questions are equivalent to $QAP$,
and therefore also $\NQP$-complete.

Determining non-zero acceptance probability of a {\em classical\/}
machine is complete for $\NP$, but determining exact accepting
probability is much harder: it is hard for $\sharpP$.  By analogy, one
might have hoped $QAP$ would be significantly easier than the problem
of determining the exact accepting probability of a {\em quantum\/}
computation, and possibly even to locate $QAP$ within the polynomial
hierarchy.  Our work shows that this is probably not the case as if
$QAP$ is in the polynomial hierarchy then this hierarchy collapses.

Work of Bennett {\em et al.} \cite{BBBV97} and recently of Fortnow and
Rogers \cite{FR:BQP} has suggested that quantum computation with
bounded error probability ($\BQP$) is most likely unable to solve
$\NP$-hard problems.  Combined with our result, this implies that 
$\BQP$ is even less likely than $\PH$ to contain $QAP$. 
We take this as evidence that quantum computers, even if
implemented, will be unable to amplify exponentially small
probabilities to such an extent that they become reliably detectable
by means of repeated experiments and observations.
This difference between bounded error computation and determining
non-zero 
acceptence probability exists classically as well; in the classical
case, bounded error computation corresponds to $\BPP$ and determining
non-zero acceptence probability
corresponds to 
$\NP$. 

Our work is part of an on-going effort to compare the power and
limitations of quantum computers with those of more well-studied
classical computers.  In the classical case, one attempts to classify
problems according to their intrinsic computational difficulty
(complexity).  For example, the class $\P$ of problems decidable by
deterministic computations running in time bounded by a polynomial in
the size of the input (i.e., {\em polynomial time}) is widely regarded
as capturing feasible, exact computations; the class $\BPP$, defined
similarly except using probabilistic machines, captures the notion of
feasible probabilistic decidability.  

Over time, complexity theorists
have built up elaborate frameworks of classes describing the power of
various models of computation.  Of these frameworks, the best known
is the polynomial hierarchy ($\PH$), the levels of which
consist of problems definable by (a fixed constant number of)
alternating polynomially bounded versions of the
quantifiers $\exists$ and $\forall$ in front of a $\P$ predicate.  The
class $\NP$, containing the well-known $\NP$-complete problems, is
the first level of this hierarchy.  It is widely believed that
$\PH$ does not collapse, i.e.,~that it is a proper hierarchy with each
level distinct from all other levels.
This implies and generalizes the conjecture that $\P \not= \NP$.
For a good introduction
to complexity theory see, for example, Balc\'azar {\em et al.}
\cite{BDG:complexity}.

Problems related to counting, e.g., ``How many satisfying truth
assignments are there to a given Boolean formula?'', have also been
widely studied (see \cite{Schoening:counting,Fortnow:counting} for
example).  It has been found \cite{Toda:PHinBPParityP,TO:PHinBPGapP}
that there are counting problems at least as difficult as any
problem in $\PH$, and thus (likely) much more difficult than any $\NP$
problem. 

The relationship between quantum computing and counting problems has
been previously observed \cite{Simon:QuantumPower,FR:BQP,BBBV97}. 
Our result further strengthens the connections between quantum
computation and counting complexity and strengthens previous results in
this area by providing the first example of a quantum
computation problem whose complexity can be precisely characterized 
in terms of a counting class.

The essential distinction between classical probabilistic models and
quantum machines, and the true source of power in the latter, rests in
the fact that the states in a quantum superposition can cancel each
other, a phenomenon known as destructive interference.  Since many
states
can be involved in such a cancellation, certain
measurable properties of the quantum state can be very sensitive to
the {\it number} of classically accepting paths.  
Our result, while using and extending the resulting connection
between quantum computation and counting problems,
also serves to clarify it.

\section{Probabilistic and Quantum Computation}

We let $\Sigma = \{0,1\}$.  We are interested in decision problems
(languages) over $\strs$.  Of particular interest are
the language
$$
\begin{array}{rcl}
QAP & = & \{ \pair{M,x,0^t} \mid 
\mbox{$M$ encodes a quantum machine that has}\\
 & & \mbox{\phantom{\{ $\pair{M,x} \mid $} 
non-zero probability of accepting $x$ in $t$ steps} \},
\end{array}
$$
and the class $\NQP$, which will be defined at the end of this section.

We review here briefly the models of classical probabilistic
computation and quantum computation that we will employ in this paper. 
Our development is
based on Turing machines, but can just as easily be based on quantum
circuits \cite{Deutsch85}, which are polynomially equivalent to quantum
Turing machines \cite{Yao:QuantumCircuits}.
See the references for more details regarding the models used
here \cite{Simon:QuantumPower} as well as equivalent
formulations \cite{Berthiaume:RetroII}.
Those who are already familiar with Turing machine models for 
quantum computation
can skip to the definition of $\NQP$ at the end of this section.

A classical probabilistic computation can be viewed as a tree.  Each
node in the tree is labeled with a configuration (instantaneous
description of tape contents, head location and internal state) of the
Turing Machine.  Edges in the tree are labeled with
real numbers in the interval $[0,1]$, which correspond to the
probability of a transition from the parent configuration to the child
configuration.  Each level of the tree represents one time step
(hereafter referred to as a {\em step}).  Throughout this paper we
will consider only computations (both classical and quantum) for which
the depth of the tree (time) is polynomial in the length of the input.
Probabilities can be assigned to a node by multiplying the
probabilities along the path from the root to that node.  The
probability of the computation being in configuration $c$ at time $t$
is the sum of the probabilites assigned to each node at level
$t$ that has been assigned configuration $c$.

In order for such a tree to represent a probabilistic computation, it
must be constrained by {\em locality}, and {\em classical
probability}.  Locality constraints require that the probability
assigned to the edge from one node to another correspond to the action
of one step of a probabilistic Turing machine, so in particular, the 
probability (1) is non-zero only if a Turing machine could actually make 
such a transition (thus for example, the only tape cells that can 
change are the ones which were under a head in the parent configuration), 
and (2) depends only on that part of the configuration which determines
the action of the machine, and not on the rest of the configuration or
the location in the tree.
Probability constraints require that the sum of all probabilities on
any level is always 1.  It is equivalent to require that the sum of
the probabilities on the edges leaving any node equal 1.  For the
purposes of complexity considerations, it is usually sufficient to
consider probabilities from the set $\set{0,\frac{1}{2},1}$.  If one
considers the probabilistic machine to be a Markov chain, the entire
computation can be represented by a matrix which transforms vectors of
configurations into vectors of configurations, with the coefficients
corresponding to probabilities.  

The probability that a machine
accepts on input $x$ after $t$ steps is
$$
\sum_{c \in \Gamma_{acc}} \Pr[\mbox{configuration $c$ at step $t$}
\mid \mbox{configuration $c_0$ at step $0$}]
$$
where $\Gamma_{acc}$ is the set of all accepting configurations and 
$c_0$ is the initial configuration corresponding to an input $x$.
Note that the class $\NP$ can be defined in terms of probabilistic 
machines:  A language, $L$, is in $\NP$ if and only if there is 
a probabilistic machine $M$ and a polynomial $p$ such that
$$
x \in L \iff Pr[\mbox{$M$ accepts $x$ in $p(|x|)$ steps}] \not= 0
$$

A quantum computation can be similarly represented by a tree, only now
the constraints are locality and {\em quantum probability}.  In the
quantum computation, the edges are assigned algebraic 
(see Section~\ref{sec:robustness}) 
complex-valued {\em probability amplitudes}.%
The amplitude of a node is again the product
along the path to that node.  The amplitude associated with being in
configuration $c$ at step $t$ is the sum of the amplitudes of all
nodes at level $t$ labeled with $c$.  The probability is the squared
absolute value of the amplitude.  A configuration $c$ uniquely
corresponds to a quantum state, denoted by $\ket{c}$. The states
$\ket{c}$, for all configurations $c$, form an orthonormal basis in a
Hilbert space. At each step we consider a quantum computation to be in
a superposition $\ket{\varphi}$ of basis states, and write this as
$$
\sum_{c \in \Gamma} \alpha_c \ket{c}
$$
where $\alpha_c$ is the amplitude of $\ket{c}$.  Since the basis
states $\ket{c}$ are mutually orthonormal, the amplitude $\alpha_c$ of
$\ket{c}$ in a superposition $\ket{\varphi}$ is the inner product of
$\ket{c}$ with $\ket{\varphi}$, denoted by $\braket{c}{\varphi}$. The
probability of accepting is defined as for the probabilistic
computation.

Once again the sum of the probabilities on any level must be 1 ($\sum
|\alpha_c|^2 = 1$).  As before, a restricted set of amplitudes for
local transitions is sufficient, namely rational numbers or square
roots of rational numbers.  In fact, the machine we construct
will only use amplitudes in
$\{ 0, \pm\frac{1}{\sqrt{2}}, \pm 1\}$.%
 It is {\em not}, however,
sufficient to require that the sum of the squares of the amplitudes
leaving any node be 1.  This is due to the effects of {\em
interference\/} among the configurations.  A quantum computation can
also be represented by a matrix which transforms quantum states into
quantum states (represented as vectors in a Hilbert space with basis
states $\ket{c}$, i.e., states of form $\ket{\varphi}$ as above).  To
satisfy the constraints of quantum probability, this matrix must be
{\em unitary} (its inverse is its conjugate transpose).  In the case
where all amplitudes are real numbers, a
matrix is unitary if and only if it is orthogonal.

The class $\NQP$ is defined, as in \cite{ADH97},
analogously to the class $\NP$ by replacing
the probabilistic machine with a quantum machine:

\begin{mytheorem}{Definition}\label{def:NQP}
A language $L$ is in $\NQP$ if and only if there is a quantum Turing
machine $Q$ and a polynomial $p$ such that 
$$ x \in L \iff Pr[\mbox{$Q$ accepts $x$ in $p(|x|)$ steps}] \not= 0
$$
\end{mytheorem}
It is not hard to see that $QAP$ is hard for $\NQP$ via a standard
argument: given $L$, $Q$, and $p$ as in Definition~\ref{def:NQP}
above, we reduce $L$ to $QAP$ by mapping input $x$ to
$\tuple{Q,x,0^{p(|x|)}}$.  We also have $QAP \in \NQP$ as a
consequence of the construction of an efficient universal quantum
machine \cite{BV97}.  Therefore, $QAP$ is complete for $\NQP$.

One might entertain other possibilities for defining a quantum analog
of $\NP$. One justification for our definition is that $\NQP$ bears
the same relation to $\BQP$ as the class $\NP$ does to $\BPP$.  As
$\BQP$ plays a central role in efficient quantum computation, this
seems like a natural definition to study. Two other possible quantum
analogs to $\NP$ would be the class $\exists \EQP$, i.e., the class of
sets $\{S |$ there is a polynomial $p$ and an $\EQP$ machine $M$ such
that for all strings $x$, $x \in S$ iff there is a string $y$ with
$|y| \le p(|x|)$ such that $M$ accepts $\langle x,y \rangle\}$ and the
class $\exists \BQP$, defined similarly.  Each of these definitions is
analogous to that of $\NP$ as $\exists\P$.

It is not clear whether any two of the three classes $\NQP$, $\exists
\EQP$, and $\exists \BQP$ are the same.  For example, it is not known
if $\P = \EQP$, but if $\P = \EQP$ and the polynomial hierarchy
separates, then $\exists\EQP = \NP \neq \NQP$.

\section{Main Result}
\label{sec:main}

Theorem~\ref{thm:GapPtoQuantum} shows 
how to design quantum machines for which the resulting amplitude of
the unique accepting state is closely related to some given function in
the class $\GapP$.  
Before giving the proof, we define this class of functions.

\begin{mytheorem}{Definition}
Given any $L\subseteq \strs$, let
$L_x = \{ y\in\strs \mid \tuple{x,y} \in L \}$.
A function  $f : \strings \to {\bf Z}$ 
is in $\GapP$ if there is a language $L$ in $\P$ 
and an integer
$k$ such that,
$$ 
f(x) = \frac{|\Sigma^{n^k} \cap L_x| - |\Sigma^{n^k} - L_x|}{2}\; ,
$$
where $n = |x|$.
\end{mytheorem}
This is equivalent to saying that a $\GapP$ function is the difference
(gap) between the number of accepting paths and the number of rejecting
paths in some nondeterministic polynomial time computation.
More information can be found in the references \cite{FFK:gaps} 
about the intuition behind this 
definition and the basic properties of the class $\GapP$.

Now we are ready to prove the technical theorem on which
Theorem~\ref{thm:equiv} rests. This result can be obtained as a
corollary of Theorem 8.9 of Bernstein and Vazirani \cite{BV97}
regarding Fourier sampling.  Our proof, which uses the same
techniques, is more direct, and will be used to generalize a result of
Fortnow and Rogers which is proved in the appendix of this paper (see
Section~\ref{sec:robustness}).

\begin{mytheorem}{Theorem}\label{thm:NQP-GapP}
\label{thm:GapPtoQuantum}
For any $f\in\GapP$, there is a ptime quantum Turing machine $Q$ and a
polynomial
$p$ such that, for all $x$ of length $n$,
\[ \Pr[\mbox{$Q(x)$ accepts}] = \frac{f(x)^2}{2^{p(n)}}. \]
In fact, for all $x$, \ $Q(x)$ has a unique accepting configuration
which it reaches with probability amplitude exactly
$-f(x)/2^{p(n)/2}$.
\end{mytheorem}

\noindent{\bf Proof Sketch: } Our proof directly uses techniques of Simon
\cite{Simon:QuantumPower} 
and Deutsch and Jozsa \cite{DeutschJozsa}.  Let $k\in\nums$
and let $L\subseteq\strs$ be a set in $\P$ such that for all $x$ of length
$n$,
\[ 
f(x) = \frac{|\Sigma^{n^k} \cap L_x| - |\Sigma^{n^k} - L_x|}{2}.
\]
Let $M$ be a polynomial time machine recognizing $L$, so that for all
$\tuple{x,y}$, $\tuple{x,y} \in L$ iff $M$ accepts on input
$\tuple{x,y}$. Fix an
input $x$ of length $n$ and let $m = n^k$.  When our quantum machine $Q$ takes
$x$ on its read-only input tape, it will use $m+1$ bits of a special work tape
$t$.  It will use other work tapes only for deterministic, reversible
computation.  We denote a possible configuration of $Q(x)$ as a basic state
\[ 
\ket{x,\bfvec{y},b} 
\]
where $x$ is the contents of the
input tape and $\bfvec{y},b$ are the contents of $t$ ($\bfvec{y}$ is a
vector of $m$ bits, and $b$ is a single bit).  We suppress the other
configuration
information, i.e., the state of $Q$, the positions of the heads, and
the contents of the other work tapes.  This other information is
irrelevant because at all important steps of the computation,
the same state and head positions of $Q$ will appear in all
configurations in the superposition, and all other work tapes besides
$t$ will be empty.  

Initially, $\bfvec{y} = \bfvec{0}$ and $b = 0$.  $Q$ first scans over
all the bits of $\bfvec{y}$ and applies to each bit what has become a useful
and popular local transition rule
\begin{eqnarray*}
\ket{0} & \mapsto & \frac{1}{\sqrt{2}}(\ket{0} + \ket{1}) \\
\ket{1} & \mapsto & \frac{1}{\sqrt{2}}(\ket{0} - \ket{1}).
\end{eqnarray*}
In general, scanning an arbitrary state
$\ket{x,\bfvec{y},b}$ in this way yields
\[ \ket{x,\bfvec{y},b} \mapsto \frac{1}{2^{m/2}}
\sum_{\bfvec{y}'} (-1)^{\bfvec{y}\cdot\bfvec{y}'} \ket{x,\bfvec{y}',b}, \]
where $\bfvec{y}\cdot \bfvec{y}'$ is the dot product $\sum_{i=1}^m y_i y'_i$ of
the bit vectors $\bfvec{y}$ and $\bfvec{y}'$. The
above transformation \cite{DeutschJozsa,Simon:QuantumPower} is called the Fourier
transform of the basis $\ket{x,\bfvec{y},b}$.  Thus $Q$ scanning the
first $m$ bits of the tape $t$ corresponds to the global transition \[
\ket{x,\bfvec{0},0} \mapsto \frac{1}{2^{m/2}} \sum_{\bfvec{y}}
\ket{x,\bfvec{y},0}. \]

$Q$ then simulates the deterministic
computation of $M$ on input $\tuple{x,\bfvec{y}}$ in a reversible
manner \cite{Deutsch85,Benioff82}, using other work tapes\footnote{This
computation is also done obliviously so that the internal state and tape head
position of the machine is the same for all components of the superposition at
any given time.  If we had used quantum circuits for the proof, this
technicality  would have been unnecessary.}.
Let $b_{\bfvec{y}}$ be the one-bit
result of the computation
of $M(x,\bfvec{y})$.  $Q$ sets $b = b_{\bfvec{y}}$.
The superposition is now
\[ \frac{1}{2^{m/2}} \sum_{\bfvec{y}} \ket{x,\bfvec{y},b_{\bfvec{y}}}. \]
Afterwards, $Q$ repeats the scan it performed at the beginning, using
the same local transformation rule, except that it now includes all $m+1$
bits, including $b$, in the scan.  This leads $Q$ into a new superposition
\[ \ket{\psi} = \frac{1}{\sqrt{2}} \frac{1}{2^m}
\sum_{\bfvec{y}} \sum_{\bfvec{y}',b'}
(-1)^{\bfvec{y}\cdot \bfvec{y}' + b_{\bfvec{y}}b'} \ket{x,\bfvec{y}',b'}. \]

We now consider the coefficient of $\ket{x,\bfvec{0},1}$ in
$\ket{\psi}$:  
\begin{eqnarray*}
\braket{x,\bfvec{0},1}{\psi}
& = & \frac{1}{\sqrt{2}} \frac{1}{2^m}
\sum_{\bfvec{y}}
(-1)^{\bfvec{y}\cdot \bfvec{0} + b_{\bfvec{y}}1} \\
& = & \frac{1}{\sqrt{2}} \frac{1}{2^m}
\sum_{\bfvec{y}}
(-1)^{b_{\bfvec{y}}} \\
& = & - \frac{1}{\sqrt{2}} \frac{1}{2^{m-1}} f(x).
\end{eqnarray*}
Finally, $Q$ deterministically looks at the $m+1$ bits of the
tape $t$.  If it sees $\bfvec{0},1$ it accepts; otherwise, it rejects.

Thus $\ket{x,\bfvec{0},1}$ is the unique accepting configuration of
$Q$, and it has probability amplitude
\[ - \frac{1}{\sqrt{2}} \frac{1}{2^{m-1}} f(x) \]
which implies the theorem by setting $p(n) = 2m-1 = 2n^k - 1$. \QED

\bigskip

A converse to Theorem~\ref{thm:NQP-GapP} follows directly from work of
Fortnow and Rogers \cite{FR:BQP}. Fortnow and Rogers' result is given
only for quantum machines that use rational amplitudes. Their proof
can be easily modified to obtain the following. In
Section~\ref{sec:robustness} we also give a generalization of this
theorem to arbitrary algebraic amplitudes.

\begin{mytheorem}{Theorem}[Fortnow, Rogers]\label{thm:GapP-NQP}
\label{thm:QuantumtoGapP}
For any ptime quantum machine $M$ 
(with transition amplitudes that are products of rational numbers and
the square root of a fixed integer), there is a $\GapP$ function
$f$, a natural number $d$, and a polynomial $p$ such that $M$ accepts any
input $x$ with probability exactly $f(x)/d^{p(|x|)}$.
\end{mytheorem}

Combining Theorems~\ref{thm:GapPtoQuantum} and \ref{thm:QuantumtoGapP} 
provides an exact characterization of $\NQP$
in terms of a counting class known to be hard for $\PH$.

\begin{mytheorem}{Definition}\label{def:C=P}
  A language $L$ is said to be in the class $\CequalsP$ if there is a
$\GapP$ function $f$ such that for any $x$, $x \in L$ if and only if 
$f(x) = 0$.
The class $\coCequalsP$ is the set of all languages with complements in
$\CequalsP$.
\end{mytheorem}

  By Theorems~\ref{thm:GapPtoQuantum} and \ref{thm:QuantumtoGapP},
a language $L$ is in $\CequalsP$ (resp., $\coCequalsP$) if and only if
there is a polynomial-time quantum Turing machine $Q$ such that for any $x$, 
$$
x \in L
\iff  \Pr[\mbox{$Q(x)$ accepts}] = 0 \;\;
\mbox{(resp., $\Pr[\mbox{$Q(x)$ accepts}]  \not = 0$)}.
$$
Thus $\NQP = \coCequalsP$, so Theorem \ref{thm:equiv} is a corollary
of Theorems~\ref{thm:GapPtoQuantum} and \ref{thm:QuantumtoGapP}.

It is known that $\CequalsP$ is hard for the polynomial hierarchy
under randomized reductions~\cite{TO:PHinBPGapP,Tarui93}. 
Thus Corollary~\ref{thm:hard} ($QAP$ is hard for $\PH$ under randomized
reductions) follows.

Hence if $QAP$ is anywhere in $\PH$, then $\PH$ collapses; in fact,
the counting hierarchy also collapses.\footnote{This is a hierarchy
built over the class $\PP$ instead of $\NP$.  The counting hierarchy
was originally defined in terms of counting quantifiers
\cite{Wagner:C=P}. The assertion follows from an alternative
characterization in terms of oracles \cite{Toran:counting}.}
Combining our results with those of Fortnow and Rogers \cite{FR:BQP},
we find that $QAP \in \BQP$ (or $QAP \in \EQP$) also implies the
collapse of the counting hierarchy.

\section{Robustness of $\NQP$}\label{sec:robustness}

In our definition of $\NQP$ we assume that the probability amplitudes are
algebraic.  In this section we want to
explore briefly the extent to which this assumption is significant.
Let $\NQP_S$
be the class defined like $\NQP$, but with amplitudes taken from the
set $S$.  So $\NQP = \NQP_\algebraics$, where $\algebraics$ is the
algebraic complex numbers.
Similar notation applies to other quantum classes.

Adleman et al.~\cite{ADH97} show that, although $\BQP_{\complexes}$ is
uncountable, $\BQP_\algebraics = \BQP_\rats = \BQP_{\set{0, \pm 3/5,
\pm 4/5, \pm 1}}$, and so the latter class provides a reasonable, robust
definition for $\BQP$. 
The proof of Theorem~\ref{thm:NQP-GapP} shows that 
$\coCequalsP \subseteq \NQP_{\set{0, \pm \frac{1}{\sqrt{2}}, \pm 1}}$, 
and it can be modified to show that 
$\coCequalsP \subseteq \NQP_{\set{0, \pm 3/5, \pm 4/5, \pm 1}}$ as well.
A proof of this modified result is given in the appendix.
These inclusions together with Corollary~\ref{cor:robust} below 
show that 
$\NQP_\algebraics = \NQP_{\set{0, \pm 3/5, \pm 4/5, \pm 1}} =
\coCequalsP$, generalizing a theorem of Fortnow-Rogers (Theorem~\ref
{thm:QuantumtoGapP}).

We use the following theorem, the main theorem for this section, which
unifies and generalizes some of the results of Adleman, {\em et al.\/}
\cite{ADH97} given above.  Our proof is somewhat similar to theirs.
We begin by recalling some basic facts from algebra.  Let
$\alpha_1,\ldots,\alpha_n$ be complex numbers.  Let
$\rats(\alpha_1,\ldots,\alpha_n)$ be the smallest subfield of
$\complexes$ containing $\alpha_1,\ldots,\alpha_n$. A basic fact of
abstract algebra is that $\alpha_1,\ldots,\alpha_n$ are all algebraic
(over $\rats$) iff $\rats(\alpha_1,\ldots,\alpha_n)$ (as a vector
space over $\rats$) is finite dimensional.

\begin{mytheorem}{theorem}
\label{thm:algebraic}
Let $M$ be any quantum accept/reject TM that has algrebraic transition
amplitudes and runs in time $t(n)$.  Then there are positive
integers $s$ and $D$, real algebraic
numbers $\alpha_1,\ldots,\alpha_s$ linearly independent over
$\rats$, and $\GapP$ functions $f_1,\ldots,f_s$, such that for
any input $x$ of length $n$,
\[ \Pr[\mbox{$M(x)$ accepts}] = \frac{1}{D^{t(n)}} \sum_{j=1}^s
f_j(x,0^{t(n)}) \alpha_j.\]
Moreover, all the $\alpha_j$ are in the field extension of $\rats$
generated by the transition amplitudes of $M$.
\end{mytheorem}

\noindent{\bf Proof Sketch:}
The transition amplitudes mentioned
in $M$ (not necessarily real), together with their
complex conjugates, generate a field $F$ that has finite dimension
over $\rats$ and that is closed under complex conjugate.  Let
$\beta_1,\ldots,\beta_m$ be a basis for $F$.  Every element of $F$ can
be expressed uniquely as a linear combination of the $\beta_i$.
Furthermore, there are unique rationals $\{q_{i,j,k}\}_{1\leq i,j,k
\leq m}$ such that $\beta_i\beta_j = \sum_k
q_{i,j,k}\beta_k$. Hence for any two elements $a = \sum a_i\beta_i$ and $b =
\sum b_i\beta_i$ of $F$, the coefficient of $\beta_k$ in $ab$ is
$\sum_i \sum_j a_i b_j q_{i,j,k}$.  Now choose
$\alpha_1,\ldots,\alpha_s$ to be a basis of $F\intersect\reals$
over $\rats$ such that for each $i$ we can write
\[ \Re(\beta_i) = \sum_{j=1}^s c_{i,j}\alpha_j, \]
where the $c_{i,j}$ are all integers.

We may assume WLOG that the $q_{i,j,k}$ are all integers.  
If not, we redefine the basis to clear all the denominators: let
$\ell$ be the lcm of all denominators appearing in the $q_{i,j,k}$.  Then
redefine the $\beta_i$ by $\beta'_i = \ell\beta_i$.
Then, $\beta'_i\beta'_j = \sum_k \ell q_{i,j,k}\beta'_k$,
so the coefficients are now all integers.

Fix any input $x$, and let $U$ be the global
unitary 1-step
transition matrix for $M(x)$.  It
is clear that each entry of $U$ is in $F$, and moreover there is an
integer $d$ and an integer-valued $\FP$ function $u$ such that the
$(i,j)$th entry of $U$ is
\[ (U)_{i,j} = \frac{1}{d} \sum_{k=1}^m u(x,i,j,k)\beta_k. \]

The proof now proceeds as in the proof of lemma 3.2 of Fortnow and
Rogers \cite{FR:BQP}, except that here we add and multiply elements of
$F$. Multiplying $U$ times itself then reduces to obtaining uniform
exponential sums of polynomial products of the $u(x,i,j,k)$'s and
$q_{i,j,k}$.
But $\GapP$ is closed under these operations.  So there
are $\GapP$ functions $g_1,\ldots,g_m$ such that the $(i,j)$th entry
of $U^t$ is
\[ (U^t)_{i,j} = \frac{1}{d^t} \sum_{k=1}^m g_k(x,0^t,i,j,k)\beta_k. \]
Now take $t = t(n)$ to be the running time of $M(x)$.  The
acceptance probability is the sum of squared absolute values of all
``accepting'' entries of $U^tS$, where $S$ is the column vector representing
the basic quantum state of the initial configuration of $M(x)$.  (Note that
squaring an absolute value is just a field operation in $F$, since $F$
is closed under complex conjugate.)  Again using
the closure properties of
$\GapP$, there are $\GapP$ functions $h_1,\ldots,h_m$ such that
\[ \Pr[\mbox{$M(x)$ accepts}] = \frac{1}{D^t} \sum_{i=1}^m
h_i(x,0^t)\beta_i, \]
where $D = d^2$.  Since this quantity is real, we have
\begin{eqnarray*}
\Pr[\mbox{$M(x)$ accepts}]
& = & \frac{1}{D^t} \sum_{i=1}^m h_i(x,0^t)\Re(\beta_i) \\
& = & \frac{1}{D^t} \sum_{i=1}^m h_i(x,0^t) \sum_{j=1}^s c_{i,j}
\alpha_j \\
& = & \frac{1}{D^t} \sum_{j=1}^s f_j(x,0^t) \alpha_j
\end{eqnarray*}
where for each $j$, we define
\[ f_j(x,0^t) = \sum_{i=1}^m c_{i,j} h_i(x,0^t). \]
It follows from the closure properties of $\GapP$ that the $f_j$ are
all in $\GapP$.  This proves the theorem.
\QED

\begin{mytheorem}{cor}[implicit in \cite{ADH97}]
\label{cor:robust}
For $M$ as above, the set 
$$
\set{x \mid \Pr[\mbox{$M(x)$ accepts}] = 0}
$$ 
is in $\CequalsP$.  Thus $\NQP_{\algebraics} \subseteq
\coCequalsP$.
\end{mytheorem}

\noindent{\bf Proof.}
Since the $\alpha_j$ are all linearly independent over $\rats$,
the probability is zero iff all the $f_j(x)$ are zero, iff $f(x) = 0$
where
\[ f(x) = \sum_{j=1}^n [f_j(x)]^2. \]
The function $f$ is clearly in $\GapP$.
\QED

The proof of Theorem~\ref{thm:algebraic} actually yields a more
general result regarding probability amplitudes, which may be of
independent interest.  As with Adleman {\em et al.}, we simply choose
a single primitive element for the field extension of $\rats$
generated by the transition amplitudes of the machine in question.

\begin{mytheorem}{theorem}
Let $Q$ be any quantum TM whose transition amplitudes are all algebraic
numbers.  There exists an algebraic number $\beta$, positive integers $d$
and $k$, and $\GapP$ functions $f_i(x,u,s)$ for all $i$,\ $0\leq i < k$
such that, for any input $x$, time $t\in\nums$, and basis state
$|s\rangle$ of $Q(x)$, the probability amplitude of $|s\rangle$ in the
quantum state of $Q(x)$ after running $t$ steps is exactly
\[ \frac{1}{d^t} \sum_{i=0}^{k-1} f_i(x,0^t,s)\beta^i. \]
Furthermore, $\beta$ is a primitive element with degree $k$ of the field
extension of $\rats$ generated by the transition amplitudes of $Q$.
\end{mytheorem}

\section{Conclusion}

  One may ask if a polynomial-time probabilistic Turing machine has a
non-zero acceptance probability. This problem is $\NP$-complete. 
$QAP$ is the analogous problem in the quantum setting
and it is $\NQP$-complete.
As we have seen in this paper, 
$\NQP = \coCequalsP$, which is a much harder class than $\NP$, and our
characterization shows that $QAP$ is nowhere in the polynomial hierarchy
unless the polynomial hierarchy and the counting hierarchy collapse and are 
equal.

We interpret this as a lower bound on the capabilities of quantum computers.
Just as it is unlikely that an $\NP$ machine's acceptance probability can be
amplified (i.e., that $\NP \subseteq \BPP$), so is it unlikely that a
quantum machine's acceptance probability can be amplified 
(i.e., $\coCequalsP \subseteq \BQP$), 
and even more unlikely that it can be amplified classically
(i.e., $\coCequalsP \subseteq \BPP$). To our knowledge, this is the
first hardness result of this nature regarding quantum computation. 
The result also shows
how destructive interference can lead to vastly different behaviors for
acceptance probabilities in classical and quantum machines.

Note that the results here show that if 
$\NQP \subseteq \BQP$, then the counting hierarchy
collapses to $\PP$. It would be interesting to see if it collapses even
farther (say, to $\BQP$). 
This would give us a better understanding of how much harder
$\NQP$ is than $\BQP$.

\section*{Acknowledgements}
The work of S. Fenner was supported in part by the NSF under grant
NSF-CCR-95-01794. The work of S. Homer was supported in part by the NSF under grant
NSF-CCR-94-00229. The work of R. Pruim was done while visiting the Computer Science
Department at Boston University. We thank C. Pollett, J.
Watrous and the referees for helpful comments.

An earlier version of this paper appeared in the 
Sixth Italian  Conference on Theoretical Computer Science,
October, 1998 \cite{FGHP:thispaper}.

\appendix

\section{Appendix}

  In this appendix, we show that Theorem~\ref{thm:NQP-GapP} also
holds for quantum machines that use amplitudes in the set
$R = \set{0, \pm \frac{3}{5}, \pm\frac{4}{5}, \pm 1}$.  This result
was first suggested to us by J.~Watrous \cite{Watrous:rat}.

\begin{mytheorem}{theorem}\label{thm:rat}
For any $f\in\GapP$, there is a ptime quantum Turing machine $Q$ with
transition amplitudes in $R$ and a polynomial
$p$ such that, for all $x$ of length $n$,
\[ 
\Pr[\mbox{$Q(x)$ accepts}] = (\frac{12}{25})^{p(n)} f(x)^2 \; . \]
In fact, for all $x$,  $Q(x)$ has a unique accepting configuration
which it reaches with probability amplitude exactly
$ (\frac{12}{25})^{p(n)/2} f(x)$.
\end{mytheorem}

\noindent{\bf Proof Sketch: }
  We indicate the essential differences with the proof of
Theorem~\ref{thm:NQP-GapP}.
  Now the basic states are $\ket{x,\bfvec{y},b}$
where $x$ and $\bfvec{y}$ are as before and $b$ is {\it two} bits.

Initially, $\bfvec{y} = \bfvec{0}$ and $b = 00$.
In $Q$'s initial scan over the bits of $\bfvec{y}$, apply the
 following local transition rule
$A$ to each bit:
\begin{eqnarray*}
\ket{0} & \mapsto & \frac{1}{5}(3\ket{0} + 4\ket{1}) \\
\ket{1} & \mapsto & \frac{1}{5}(-4\ket{0} + 3\ket{1}).
\end{eqnarray*}

If we let $\Count{i}{\bfvec{z}}$ be the number of components of the vector 
$\bfvec{z}$ that have the value $i$, then 
transforming an arbitrary state $\ket{x,\bfvec{y},b}$ in this way yields
\[ 
\ket{x,\bfvec{y},b} \mapsto 
\frac{1}{5^{m}}
\sum_{\bfvec{y}'} 
(-4)^{\Count{-1}{\bfvec{y}' - \bfvec{y}} }
 4^{\Count{1}{\bfvec{y}' - \bfvec{y}} } 
 3^{\Count{0}{\bfvec{y}' - \bfvec{y}} } 
\; \ket{x,\bfvec{y}',b} \; . 
\]
Thus $Q$ scanning the
first $m$ bits of the tape corresponds to the global transition 
\[
\ket{x,\bfvec{0},00} \mapsto 
\sum_{\bfvec{y}}
\alpha_{\bfvec{y}} 
\ket{x,\bfvec{y},00}, 
\]
where 
$$
\alpha_{\bfvec{y}} = 
\frac{
       3^{\Count{0}{\bfvec{y}}}  4^{\Count{1}{\bfvec{y}}}
     }
     {5^{m}} \; .
$$

Again $Q$ simulates the deterministic
computation of $M$ on input $\tuple{x,\bfvec{y}}$ in a reversible
manner.
Let $b_{\bfvec{y}}$ be $01$ if $M$ accepts, $10$ if it rejects.
$Q$ sets $b = b_{\bfvec{y}}$.

The superposition is now
\[
\ket{x,\bfvec{0},00} \mapsto 
\sum_{\bfvec{y}}
\alpha_{\bfvec{y}} 
\ket{x,\bfvec{y},\\b_{\bfvec{y}}}, 
\]
Next, $Q$ performs the transition $A$ to the first bit of 
$b_{\bfvec{y}}$ and then  the transition $B$ given by

\begin{eqnarray*}
\ket{0} & \mapsto & \frac{1}{5}(4\ket{0} + 3\ket{1}) \\
\ket{1} & \mapsto & \frac{1}{5}(-3\ket{0} + 4\ket{1}).
\end{eqnarray*}
to the second bit of $b_{\bfvec{y}}$.  That is, $Q$ applies
$$
T = 
\frac{1}{25} \left[ 
\matrix{ 12 & 9 & 16 & 12 \cr -9 & 12 & -12 & 16 \cr -16 & -12 & 12 & 9 \cr
   12 & -16 & -9 & 12 \cr  }
\right]
$$
to $b_{\bfvec{y}}$.
Finally,
$Q$ repeats the scan it performed at the beginning, using
the same local transformation rule ($A$) on the $m$ bits of $\bfvec{y}$.
This leads $Q$ into a new superposition
\[ 
\ket{\psi} = \frac{1}{5^{2m+2}}
\sum_{\bfvec{y}} \sum_{\bfvec{y}',b'}
\alpha(\bfvec{y},\bfvec{y}') \;
\beta(b_{\bfvec{y}},b') \;
\ket{x,\bfvec{y}',b'},  
\]
where 
\begin{eqnarray*}
\alpha(\bfvec{y}, \bfvec{y}') & = &
(-4)^{\Count{-1}{\bfvec{y}' - \bfvec{y}} + \Count{-1}{\bfvec{y} - \bfvec{0}}}
 4^{\Count{1}{\bfvec{y}' - \bfvec{y}}  + \Count{1}{\bfvec{y} - \bfvec{0} }}
 3^{\Count{0}{\bfvec{y}' - \bfvec{y}}  + \Count{0}{\bfvec{y} - \bfvec{0} }},
\\
\beta(\bfvec{y}, \bfvec{y}') & = & 25 \cdot T_{\bfvec{y},\bfvec{y}'} \; .
\end{eqnarray*}
We now consider the coefficient of $\ket{x,\bfvec{1},01}$ in
$\ket{\psi}$:  
\begin{eqnarray*}
\braket{x,\bfvec{0},1}{\psi}
& = & 
\sum_{\bfvec{y}}
 4^{\Count{1}{\bfvec{1} - \bfvec{y}}  + \Count{1}{\bfvec{y} - \bfvec{0} }}
 3^{\Count{0}{\bfvec{1} - \bfvec{y}}  + \Count{0}{\bfvec{y} - \bfvec{0} }}
\beta(b_{\bfvec{y}},01) 
\\
& = & \frac{1}{5^{2m+2}} 
\sum_{\bfvec{y}}
 12^{\Count{1}{\bfvec{1} - \bfvec{0}} }
\beta(b_{\bfvec{y}},01) 
\\
& = & \frac{12^m}{25^{m+1}}
\sum_{\bfvec{y}}
\beta(b_{\bfvec{y}},01) 
\\
& = & \frac{12^m}{25^{m+1}} 12 f(x).
\end{eqnarray*}
Finally, $Q$ deterministically looks at the $m+2$ bits of the
tape $t$.  If it sees $\bfvec{1},01$ it accepts; otherwise, it rejects.

Thus $\ket{x,\bfvec{1},01}$ is the unique accepting configuration of
$Q$, and it has probability amplitude
\[ 
(\frac{12}{25})^{m+1} f(x)\; ,
\]
which implies the theorem by setting $p(n) = 2m+2 = 2n^k + 2$. \QED

\end{document}